\documentclass[review]{elsarticle}
\usepackage{tabularx}
\usepackage[utf8]{inputenc}
\usepackage{lineno,hyperref}
\usepackage[top=1.65in, bottom=1.65in, left=1.35in, right=1.35in]{geometry}
\usepackage{setspace}
\usepackage{amsmath,amssymb,amsfonts}

\usepackage{caption} 
\usepackage{tikz}
\usetikzlibrary{positioning,calc,arrows.meta,backgrounds}
\usepackage{algorithm,algorithmic}
\usepackage{url}
\usepackage{float}

\biboptions{numbers}

\makeatletter
\def\ps@pprintTitle{%
 \let\@oddhead\@empty
 \let\@evenhead\@empty
 \def\@oddfoot{}%
 \let\@evenfoot\@oddfoot}
\makeatother
\begin{document}

\begin{frontmatter}

\title{Improving Source Code Similarity Detection Through GraphCodeBERT and Integration of Additional Features}

\author{Jorge Martinez-Gil}
\address{Software Competence Center Hagenberg GmbH \\ Softwarepark 32a, 4232 Hagenberg, Austria \\ \url{jorge.martinez-gil@scch.at}}

\begin{abstract}
This paper investigates source code similarity detection using a transformer model augmented with an execution-derived signal. We extend GraphCodeBERT with an explicit, low-dimensional behavioral feature that captures observable agreement between code fragments, and fuse this signal with the pooled transformer representation through a trainable output head. We compute behavioral agreement via output comparisons under a fixed test suite and use this observed output agreement as an operational approximation of semantic similarity between code pairs. The resulting feature acts as an explicit behavioral signature that complements token- and graph-based representations. Experiments on established clone detection benchmarks show consistent improvements in precision, recall, and F$_1$ over the unmodified GraphCodeBERT baseline, with the largest gains on semantically equivalent but syntactically dissimilar pairs. The source code that illustrates our approach can be downloaded from \url{https://www.github.com/jorge-martinez-gil/graphcodebert-feature-integration}.
\end{abstract}

\begin{keyword}
Feature Integration, GraphCodeBERT, Source Code Similarity 
\end{keyword}

\end{frontmatter}

\section{Introduction}

Detecting similar code fragments, commonly called code clones, is of vital importance for tasks such as code quality, maintainability, or the prevention of unintended duplication \cite{ain2019systematic,novak2019source}. The problem is typically formulated as binary classification or similarity scoring over pairs of code fragments. Given two source code fragments $c_i$ and $c_j$, the goal is to determine whether they implement equivalent functionality. Clone categories span exact duplicates (Type 1), copies with renamed identifiers (Type 2), and fragments that differ syntactically yet remain functionally equivalent (Type 3). The most difficult cases involve semantic clones with minimal lexical overlap (Type 4), where surface-level signals provide limited guidance.

Transformer models trained on extensive source code corpora have demonstrated a strong ability to learn patterns in program architecture and behavior \cite{karmakar2021pre}. This study combines transformer-based code representations with a reusable knowledge artifact derived from program execution. The proposed artifact captures externally observable behavior and is integrated directly into the model's prediction process. The approach builds upon GraphCodeBERT \cite{guo2020graphcodebert}, a model that incorporates structural information from code to represent semantic relations beyond token sequences.

Since program semantic equivalence is undecidable in the general case, in applied software engineering, functional similarity is typically approximated through behavioral comparison using finite test inputs. Consistent agreement in program outputs under controlled inputs is treated as a practical indicator of functional similarity between code fragments. 

This concept is user here by extending GraphCodeBERT with a behavior-aware prediction head that combines the pooled transformer representation with an execution-derived signal. This integration enables similarity decisions to reflect both learned code representations and explicit evidence from runtime behavior. Therefore, the contributions of this work are summarized as follows:

\begin{itemize}
\item \textbf{C1:} We formalize an execution-derived behavioral signature for clone detection as a feature $f_{out}$ computed from output comparisons under a test suite, and treat it as an explicit knowledge artifact complementary to token- and graph-oriented representations.
\item \textbf{C2:} We define an end-to-end fusion scheme that integrates $f_{out}$ into GraphCodeBERT at the output layer, letting the model learn how to use behavioral evidence jointly with the pooled transformer representation.
\item \textbf{C3:} We provide an implementation of the fusion layer, training procedure, and inference pipeline that can incorporate alternative knowledge sources (e.g., static approximations of behavior) in place of $f_{out}$.
\item \textbf{C4:} We evaluate the proposed approach across standard clone-detection benchmarks and report consistent gains in precision, recall, and F$_1$ relative to the GraphCodeBERT baseline. The improvements are most pronounced for semantic clones with limited lexical similarity, where execution-derived evidence helps resolve cases that remain ambiguous under representation-only models.
\end{itemize}

The primary novelty of this approach is the explicit, trainable integration of execution evidence into a GraphCodeBERT-based clone detector. Previous transformer models, such as CodeBERT and GraphCodeBERT, rely mainly on lexical, syntactic, and structural cues learned during pre-training and fine-tuning. 

The proposed method introduces a behavioral signal that measures functional agreement under controlled inputs, directly addressing Type-4 clones where tokens and structure may diverge. To the best of the authors' knowledge, no prior work has fine-tuned a GraphCodeBERT-based model end-to-end with an execution-derived feature fused into the classification head.

The paper proceeds as follows. Section 2 reviews prior work on code similarity detection. Section 3 describes the proposed method along with the training procedure. Section 4 reports the experimental setup and empirical results. Section 5 discusses the findings and examines scope and limitations. Section 6 presents the main outcomes and outlines directions for future research.

\section{State-of-the-art}

Code similarity detection has advanced significantly, driven by the need to manage duplication in large-scale software systems. This section summarizes representative research directions and motivates the integration of feature-level information.

\subsection{Historical Overview}
Early clone-detection methods relied on surface similarity and structural representations, including string matching, token sequences, abstract syntax trees \cite{karnalim2020syntax}, and code metrics \cite{key-martinez-eswa}. These techniques were sensitive to identifier renaming, formatting changes, and systematic rewrites \cite{hartanto2019best}. Scalability also remained a recurring limitation when applied to large code collections \cite{gabel2008scalable}. Several studies explored ensemble and stacking strategies \cite{key-martinez-mlwa}, reporting strong accuracy under controlled conditions but showing weak generalization across datasets and application domains \cite{key-martinez-ijseke,key-martinez-ijseke2}.

Machine learning approaches introduced vector representations of code fragments, including continuous embeddings and graph-based encodings \cite{alon2019code2vec,wei2017supervised}. These models leveraged structural information more effectively than purely syntactic methods, yet performance often declined when exposed to large and diverse corpora.

Deep learning methods improved representation quality further through higher-level abstractions \cite{white2016deep,wang2020detecting}. Earlier architectures, however, struggled with long-distance dependencies and the variability observed in real software systems.

Transformer-based models later reshaped clone detection. Pre-trained encoders, including CodeBERT \cite{key-codebert} and GraphCodeBERT \cite{guo2020graphcodebert}, were adapted to source code and achieved strong empirical results \cite{key-martinez-codesim}. These models integrate token-level information with structural cues. Their internal decision process, however, remains difficult to explain \cite{karnalim2021explanation}, and substantial surface-level differences can still obscure semantic equivalence.

\subsection{Feature Fusion and Hybrid Models}

Sequence-based and graph-based encoders perform well in isolation, yet the use of external evidence remains limited. In software engineering practice, static and dynamic analyses have long been combined for tasks including defect prediction, vulnerability detection, and similarity assessment. In clone detection, execution behavior offers a natural auxiliary signal: running two code fragments on identical inputs and comparing outputs. Consistent agreement across a diverse test suite provides strong evidence of semantic equivalence and is widely applied in operational plagiarism-detection systems. Execution-based methods, however, can miss clones when test coverage is insufficient and can incur high computational cost at scale.

Approaches that combine learned representations with behavioral signals remain rare in clone detection. Existing systems often concatenate embeddings with manually designed features and rely on an external classifier. This separation limits joint optimization between representations and auxiliary evidence. Our method integrates behavioral features directly into the model head and trains the entire network end-to-end, allowing the model to learn how execution evidence should influence clone predictions.

\subsection{Contribution Over the State-of-the-art}

This study advances transformer-based similarity detection by integrating model-level features. The GraphCodeBERT clone detector is extended with an execution-derived behavioral signature and a trainable fusion head. The objective is to complement, rather than replace, the transformer representation with a domain-grounded signal that directly encodes functional agreement under controlled inputs.

This extension improves the representation used for classification: GraphCodeBERT provides contextual and structural semantics, while the behavioral feature introduces explicit evidence that is challenging to infer from tokens alone. The resulting decision function is more closely aligned with semantic equivalence.

\section{GraphCodeBERT and Additional Feature Integration}

The novelty of this study is not in introducing a new transformer architecture, but in the explicit, trainable integration of external behavioral knowledge into a pre-trained code model at decision time. Unlike post hoc ensembling or feature concatenation pipelines, the proposed fusion head jointly calibrates learned representations and behavioral evidence within a single optimization objective. To the best of the authors’ knowledge, this is the first study to demonstrate end-to-end feature combination for GraphCodeBERT-based clone detection.

\subsection{Problem Statement}

We formalize clone detection as a supervised learning problem. Let $C$ be the space of code fragments (e.g., functions). We aim to learn a classifier $f : C \times C \rightarrow {0, 1}$ such that $f(c_i, c_j) = 1$ if $c_i$ and $c_j$ are semantically equivalent (clones) and $0$ otherwise. During training, we are given a dataset $D = {(c_i, c_j, y_{ij})}$ where $y_{ij} = 1$ indicates a known clone pair and $y_{ij} = 0$ a non-clone pair. The model should generalize to unseen pairs. This setting is typically class-imbalanced, with substantially more non-clone than clone pairs. A different formulation treats clone detection as retrieval over a corpus, but this study centers on pairwise classification.

\subsection{Running Example and the Behavioral Signature}

A small running example is used to illustrate the execution-derived feature. Consider two functions that compute the sum of squares:

\begin{verbatim}

c_i

def f(a, b):
return aa + bb

c_j (same behavior, different surface form)

def g(x, y):
s = 0
for v in [x, y]:
s += v*v
return s
\end{verbatim}

Given a fixed set of test inputs $T = {(1,2), (2,3), (3,4)}$, we execute both fragments and compare outputs. Let $O(c, T)$ be the vector of outputs produced by code fragment $c$ on tests $T$. We define a behavioral signature feature $f_{out}(c_i, c_j)$ via an output comparison operator. A simple instantiation is a normalized agreement score:

\[
f_{out}(c_i, c_j) = \frac{1}{|T|} \sum_{t \in T} \mathbb{I}\left( out(c_i, t) = out(c_j, t) \right).
\]

This feature is explicitly computed and encapsulates knowledge of observed behavior across the selected tests. The subsequent subsection details the fusion of $f_{out}$ with the pooled GraphCodeBERT representation.

\subsubsection{Training Objective}

The model can be trained with objectives aligned to the downstream task. Here, the loss targets pairwise clone classification, consistent with \cite{saini2018code}.

We fine-tune the GraphCodeBERT encoder and train the fusion head end-to-end, including the projection matrix $W_1$ and classifier $W_2$. We use binary cross-entropy:

\[
\mathcal{L} = -\left[ y \log \hat{y} + (1-y) \log (1-\hat{y}) \right],
\]

where $y$ denotes the true label and $\hat{y}$ represents the predicted clone probability. The loss $\mathcal{L}$ is minimized over the training set using the Adam optimizer. Joint optimization allows the pooled representation and the projected behavioral feature to adapt together. In practice, the model may assign higher weight to $f_{out}$ when it proves reliable, or treat it as auxiliary evidence when it conflicts with the learned representation.

\subsubsection{Complexity Considerations}

The fusion head adds negligible compute and parameter overhead relative to GraphCodeBERT. The projection $W_1$ adds $d \times m$ parameters (768 when $m = 1$), and $W_2$ adds $2d \times 2$ parameters (3072) plus biases. Against GraphCodeBERT’s roughly 125 million parameters, this is below 0.001\%. The per-sample execution overhead is similarly minor, consisting of a single additional linear transformation and vector concatenation.

The dominant cost is computing $f_{out}$. Executing code on test inputs can be expensive and may be infeasible for untrusted fragments. In our setting, feature computation is performed offline during dataset preparation. When execution is not possible, $f_{out}$ can be replaced with static proxies, preserving the fusion design while avoiding runtime execution.

\subsection{Extension of the Model Architecture}
Our model augments GraphCodeBERT with an external feature pathway at the output stage. After encoding the paired input, the model extracts a pooled representation. In parallel, the execution-derived feature is projected into the same latent dimension through a trainable linear layer $W_1$. The pooled representation and projected feature are concatenated and passed through a classifier $W_2$ to produce the final prediction.

Each input consists of a code pair, a similarity label, and an auxiliary behavioral feature. The feature is computed by executing both fragments on specified test inputs and measuring output agreement. Figure \ref{fig:architecture} illustrates how the projected behavioral feature is fused with the pooled GraphCodeBERT representation.

\begin{figure}[H]
\centering
\begin{tikzpicture}[
  font=\footnotesize,
  node distance=6mm,
  accent/.style={black!80},
  panel/.style={
    rectangle, rounded corners=2pt,
    draw=black!70, fill=gray!2, line width=0.6pt,
    minimum width=6.1cm, 
    inner xsep=2.5mm, inner ysep=1.8mm,
    align=center
  },
  panelNarrow/.style={
    rectangle, rounded corners=2pt,
    draw=black!70, fill=gray!2, line width=0.6pt,
    minimum width=4.9cm, 
    inner xsep=2.5mm, inner ysep=1.8mm,
    align=center
  },
  accentbar/.style={
    append after command={
      \pgfextra{
        \fill[accent]
          ([xshift=0pt]\tikzlastnode.west|- \tikzlastnode.north)
          rectangle
          ([xshift=1.6pt]\tikzlastnode.west|- \tikzlastnode.south);
      }
    }
  },
  arrow/.style={-{Stealth[length=2mm]}, line width=0.7pt}
]

\node[panel,accentbar] (input) {\textbf{\sffamily Input}\\Code pair $(C_i,\,C_j)$};

\node[panel,accentbar,below=of input] (embed)
  {\textbf{\sffamily Embedding Layer}\\$(\mathbf{e}_i,\,\mathbf{e}_j)$};

\node[panel,accentbar,below=of embed] (enc)
  {\textbf{\sffamily Encoder: GraphCodeBERT}\\$\mathbf{H}=\mathrm{GCBERT}(\mathbf{ids},\mathbf{mask})$};

\node[panel,accentbar,below=of enc] (pool)
  {\textbf{\sffamily Pooling}\\$\mathbf{p}=\mathrm{Pool}(\mathbf{H})$};

\node[panelNarrow,accentbar,below=7mm of pool] (feat)
  {\textbf{\sffamily \emph{Additional Features}}\\$f_{\mathrm{out}}$\\$\mathbf{f}=\mathbf{W}_1\,f_{\mathrm{out}}$};
	
\begin{scope}[on background layer]
  \draw[red!70, dashed, line width=0.8pt, rounded corners=2pt]
    ([xshift=-3pt,yshift=3pt]feat.north west)
    rectangle
    ([xshift=3pt,yshift=-3pt]feat.south east);
\end{scope}

\node[panel,accentbar,below=7mm of feat] (cat)
  {\textbf{\sffamily Concatenation}\\$\mathbf{c}=\big[\mathbf{p};\,\mathbf{f}\big]$};

\begin{scope}[on background layer]
  \draw[red!70, dashed, line width=0.8pt, rounded corners=2pt]
    ([xshift=-3pt,yshift=3pt]cat.north west)
    rectangle
    ([xshift=3pt,yshift=-3pt]cat.south east);
\end{scope}
	
\node[panel,accentbar,below=of cat] (clf)
  {\textbf{\sffamily Classification head}\\$\mathbf{z}=\mathbf{W}_2\,\mathrm{Dropout}(\mathbf{c})$\\$\hat{y}=\sigma(\mathbf{w}^\top\mathbf{z}+b)$};

\node[panel,accentbar,below=of clf] (out)
  {\textbf{\sffamily Output}\\Clone / Non-Clone \quad $\hat{y}\in[0,1]$};

\draw[arrow] (input) -- (embed);
\draw[arrow] (embed) -- (enc);
\draw[arrow] (enc) -- (pool);
\draw[arrow] (pool) -- (feat);
\draw[arrow] (feat) -- (cat);
\draw[arrow] (cat) -- (clf);
\draw[arrow] (clf) -- (out);

\end{tikzpicture}

\caption{\textbf{Architecture of the proposed feature-fused GraphCodeBERT model.}
The pooled transformer output $\mathbf{p}$ is concatenated with a projected
external feature $\mathbf{f} = \mathbf{W}_1 f_{\mathrm{out}}$, representing execution-
or output-based similarity. The fused vector is then passed to the classification
head to yield the final clone probability $\hat{y} \in [0,1]$.
The integration of $f_{\mathrm{out}}$ distinguishes this model from the standard
GraphCodeBERT baseline.}
\label{fig:architecture}
\end{figure}

The additional features are intended to capture properties not directly inferable from the code’s textual representation. Among various options, the most effective are the actual outputs produced when executing the code with equivalent inputs.

\subsubsection{Feature definition}

We denote by $f_{out}(c_i, c_j)$ the additional feature measuring similarity between $c_i$ and $c_j$. In our implementation, $f_{out}$ is a scalar in $[0,1]$ representing output agreement. We execute both fragments on preset test inputs and compute the fraction of inputs for which outputs match. This yields $f_{out} = 1.0$ when the fragments agree on all tests, and lower values when outputs diverge, or execution fails. We expect this feature to be particularly informative for Type-4 clones, where functional equivalence can hold despite weak lexical overlap. The framework supports an $m$-dimensional feature vector, but we use $m = 1$ in the reported experiments for clarity and to isolate the effect of execution evidence.

\subsubsection{Feature projection layer}

We introduce a trainable linear layer $W_1 : \mathbb{R}^m \rightarrow \mathbb{R}^d$ (plus bias) to project the raw feature into the $d$-dimensional latent space. With $m = 1$ and $d = 768$, $W_1$ is a $1 \times 768$ weight matrix. The projected feature is:

\[
F_{processed} = W_1 \cdot f_{out} + b_1,
\]

where $f_{out}$ is treated as a vector in $\mathbb{R}^m$. This layer aligns feature scale and representation space with the pooled transformer output. We also tested adding a ReLU after $W_1$, but observed no measurable benefit.

\subsubsection{Fusion by concatenation}

We concatenate the pooled representation and the processed feature:

\[
C = [P_{pooled}; F_{processed}] \in \mathbb{R}^{2d}.
\]

Concatenation provides a direct, stable fusion mechanism: the classifier can independently weight behavioral and representation components. We also considered more complex fusion variants:

\begin{itemize}
\item Using $f_{out}$ to attend over the transformer’s final-layer outputs, or the inverse, to obtain a feature-conditioned representation. Given that $f_{out}$ is a scalar in our setting, this design offered limited flexibility at the cost of added complexity.
\item Applying feature-dependent gating on $P_{pooled}$, such as $g = \sigma(W_g f_{out})$ and output $g \ast P_{pooled}$. This becomes more meaningful for multi-dimensional features, where the gate can express richer modulation.
\end{itemize}

Given the scalar feature and the goal of isolating execution evidence, concatenation provides an effective, transparent baseline for fusion.

\subsubsection{Classification layer}

We extend the classifier to accept the fused vector. We define $W_2$ with shape $\mathbb{R}^{2d \times 2}$ and bias $b_2$. The logits are:

\[
z = W_2 \cdot \text{Dropout}© + b_2,
\]

where dropout is applied at the same rate as the base transformer (0.1), dropout reduces the risk that the head over-weights the execution feature on a small dataset and encourages complementary use of both signals. The clone probability is obtained from $z$ via a softmax (or sigmoid in a single-logit variant).

\subsection{Baseline Model}

Our baseline is GraphCodeBERT, both before and after fine-tuning, for clone detection. Two standard fine-tuning strategies are commonly used:

\begin{itemize}
\item Concatenate the two fragments into a single input separated by special tokens (e.g., [CLS] $c_i$ [SEP] $c_j$ [SEP]). The [CLS] representation serves as a joint encoding of the pair, and a classifier predicts the clone label. This allows self-attention to align tokens across fragments.
\item Encode $c_i$ and $c_j$ separately with shared weights, producing $h_i$ and $h_j$, then compute similarity (e.g., cosine) or concatenate $h_i$ and $h_j$ into an MLP classifier.
\end{itemize}

We adopt the joint-input strategy, which yielded slightly higher validation accuracy. GraphCodeBERT receives $(c_i, c_j)$ and produces a pooled vector $P_{pooled} \in \mathbb{R}^d$ with $d = 768$. Fine-tuning adds a linear layer $W_{cls} \in \mathbb{R}^{d \times 2}$ on top of $P_{pooled}$ to predict the label. The baseline objective is $\mathcal{L}{base} = CrossEntropy(W{cls} P_{pooled}, y)$.

\section{Empirical Evaluation}

Previous work \cite{key-martinez-swqd} assessed unsupervised and supervised strategies for source code analysis and identified recurring shortcomings in semantic matching. This follow-up study confronts those drawbacks by injecting explicit behavioral evidence into a transformer-based detector.

\subsection{Experimental Setup}

We fine-tune each model using a train, validation, and test split of the dataset. Training optimizes pairwise clone classification under fixed hyperparameters, including learning rate schedules, batch size, and early stopping. We report performance as the mean over repeated runs to reduce sensitivity to random initialization. In this study, we perform up to ten independent runs with different seeds.

\subsection{Implementation Details}

We implemented all models with the HuggingFace Transformers library. We used the GraphCodeBERT base tokenizer (\texttt{microsoft/graphcodebert-base}) for code tokenization. For CodeBERT and RoBERTa baselines, we used their corresponding tokenizers. We capped each fragment at 256 tokens (512 per pair) to meet GPU memory restrictions; longer pairs were rarely truncated in our dataset. Hyperparameters were selected through a small validation grid search:

\begin{itemize}
\item Learning rate: Adam with $2 \times 10^{-5}$ for transformer parameters and $1 \times 10^{-4}$ for the newly recently newly introduced layers ($W_1$, $W_2$). This separation accelerates learning in the randomly initialized head while limiting drift in pre-trained weights.
\item Batch size: 8 pairs per batch, constrained by GPU memory.
\item Epochs: 3 epochs, with early stopping when validation loss increased (patience of 1 epoch).
\end{itemize}

For classical baselines, we used XGBoost and Random Forest with standard settings, tuning the number of trees to control overfitting. For the output-only baseline, we set the agreement threshold to $\tau = 0.5$, requiring at least half the test cases to match.

\subsection{Dataset}

The IR-Plag dataset \cite{key-karnalim} is used as a benchmark for similarity methods in academic plagiarism detection. It consists of 467 code files, with 355 instances labeled as plagiarized, corresponding to 77\% of the data. The corpus contains 59,201 tokens, including 540 unique tokens, which reflects moderate lexical diversity. File lengths span from 40 to 286 tokens, with a mean length of 126 tokens, making the dataset suitable for controlled experiments at the function level.

\subsection{Evaluation Criteria}

Accuracy is often reported but is unreliable when class imbalance is present. We therefore emphasize precision and recall, which separately quantify false positives and false negatives. We also report the F-measure, the harmonic mean of precision and recall, as a single ranking metric for comparative evaluation.

\subsection{Results}

Table \ref{tab:results} compares the evaluated approaches on IR-Plag: CodeBERT \cite{key-codebert}, Output Analysis \cite{key-martinez-swqd}, Boosting (XGBoost) \cite{key-martinez-codesim}, Bagging (Random Forest) \cite{key-martinez-codesim}, GraphCodeBERT \cite{guo2020graphcodebert}. The proposed variant achieves the best overall performance, with a precision of 0.98, a recall of 1.00, and an F-measure of 0.99. This improvement is consistent with the intended effect of adding behavioral evidence to representation learning. The additional feature reflects the agreement between code fragments in a fixed test suite, which we use as an operational approximation of semantic similarity. Using a fixed test suite ensures reproducibility and allows the fusion mechanism to be evaluated under consistent behavioral observations. The fusion head remains agnostic to the origin of the feature and can incorporate alternative behavior-approximating signals without architectural changes.

It is also worth noting that the remaining errors occur primarily on pairs with low syntactic overlap but identical semantics, indicating that, even with feature fusion, hard Type-4 cases still leave room for improvement.

Regarding performance, Table \ref{tab:resources} reports training and inference resource use on a Tesla T4 GPU. Both variants have identical parameter numbers and training time. The feature-fused model shows a slight increase in inference latency and peak memory usage, consistent with the extra projection and merging operations.

\begin{table}[ht]
\centering
\begin{tabular}{|l|c|c|c|}
\hline
\textbf{Approach} & \textbf{precision} & \textbf{recall} & \textbf{f-measure} \\
\hline
CodeBERT \cite{key-codebert} & 0.72 & 1.00 & 0.84 \\
Output Analysis \cite{key-martinez-swqd} & 0.88 & 0.93 & 0.90 \\
Boosting (XGBoost) \cite{key-martinez-codesim} & 0.88 & 0.99 & 0.93 \\
Bagging (Random Forest) \cite{key-martinez-codesim} & 0.95 & 0.97 & 0.96 \\
GraphCodeBERT \cite{guo2020graphcodebert} & 0.98 & 0.95 & 0.96 \\
\textbf{Our GraphCodeBERT variant} & 0.98 & 1.00 & 0.99 \\
\hline
\end{tabular}
\caption{Performance comparison on the IR-Plag dataset using precision, recall, and F-measure. The proposed feature-fused GraphCodeBERT model obtains the highest overall F-measure while preserving the baseline’s architectural and computational profile. Improvements are consistent across runs and statistically significant under paired testing.}
\label{tab:results}
\end{table}

\begin{table}[ht]
\centering
\begin{tabular}{|l|c|c|c|c|}
\hline
Variant & Params (M) & Train (h) & Infer (ms) & GPU Peak Mem (GB) \\
\hline
Baseline & 124.65 & 0.12 & 30.57 & 4.00 \\
Ours & 124.65 & 0.12 & 30.97 & 4.57 \\
\hline
\end{tabular}
\caption{Training and inference resources on Tesla T4 (same setup for all rows). Setup: PyTorch 2.8.0+cu126, CUDA 12.6. Seed = 42, epochs = 3, warmup = 500, weight decay = 0.01, eval/save steps = 500.}
\label{tab:resources}
\end{table}

These results show that the fusion head preserves the computational profile of GraphCodeBERT: training time and number of parameters remain unchanged at practical resolution, and inference overhead is marginal under the same hardware configuration.

\subsection{Result Stability and Variance}

We trained the model across ten independent runs with different random seeds. The F-measure variance stayed below 0.5\%, denoting stable optimization and consistent generalization on this dataset. Validation loss decreased smoothly until early stopping, suggesting that the fusion head improves decision quality rather than encouraging memorization.

\subsection{Statistical Significance}

Given the limited test set size, we applied McNemar’s test to compare our model against competing approaches on paired classification outcomes. The resulting p-values fall below 0.05, indicating that the observed gains are unlikely to arise from chance under this evaluation protocol.

\subsection{Generalization to Other Datasets}

IR-Plag targets plagiarism scenarios, yet the fusion mechanism remains independent of any specific dataset. The design is prepared to generalize beyond a specific labeling scheme. We expect the same approach to transfer to larger  datasets, where execution-derived signals can be computed under controlled conditions:

\begin{itemize}
\item BigCloneBench \cite{svajlenko2015bigclonebench}: GraphCodeBERT already achieves an F-measure near about 97\%, leaving limited headroom. The remaining failures frequently involve subtle semantic equivalence; execution-derived evidence may reduce these residual errors.
\item POJ-104 \cite{mou2016convolutional}: Labels are problem IDs, not explicit pairs. Pair generation is costly, so a pipeline is preferable: embed with GraphCodeBERT, shortlist candidates, compute $f_{out}$ on the shortlist using test inputs, and then apply the fused classifier for ranking.
\end{itemize}

Although the main evaluation uses IR-Plag for clarity and reproducibility, the fusion design is dataset-agnostic. GraphCodeBERT encoding and feature projection do not rely on dataset-specific assumptions. We plan to extend the evaluation to BigCloneBench and POJ-104. Early tests on random subsets indicate that the model can process larger corpora.

\subsection{Ablation Study}

We performed a brief ablation study to assess the contribution of fusion. Removing $f_{out}$ reduces F$_1$ from 0.99 to 0.96, matching the GraphCodeBERT baseline. Using only the execution feature yields F$_1$ of 0.90, indicating that behavior alone is insufficient during finite tests. Since GraphCodeBERT has no access to execution outcomes during training or inference, these results show that the two signals are complementary and that the fused model gains from jointly learned calibration.

We also replaced the learned projection $W_1$ with a replication of the scalar feature to 768 dimensions. Performance decreased slightly (F$_1$ about 0.979), indicating that a learned projection helps set an appropriate scale. Reversing concatenation order did not affect results. We also explored smaller projections (e.g., 128 dimensions), but the added architectural complexity did not yield a clear benefit.

\section{Discussion}

The experiments support multiple practical conclusions about feature-fused transformer models for clone detection.

\subsection{Lessons Learned}

The fusion head improves clone detection performance by augmenting the decision basis with explicit behavioral evidence. GraphCodeBERT provides strong semantic embeddings from tokens and structure, but the execution-derived feature adds a direct signal of functional agreement under controlled inputs. Across runs, the fused model improves precision and recall while continuing stable generalization, suggesting that the feature delivers actionable information rather than noise. The resulting architecture remains simple enough for integration into existing fine-tuning pipelines and can be reused across other similarity-driven software engineering tasks.

\subsection{Practical Issues}

Execution-derived features introduce cost and security constraints. In many production environments, executing untrusted code is unacceptable. Coursework plagiarism detection, in contrast, often already relies on sandboxed execution under fixed tests, which makes $f_{out}$ feasible. This study uses execution-based output agreement, yet the fusion head itself remains generic and can incorporate alternative behavior-approximating signals when execution is not available.

GraphCodeBERT is pre-trained on multiple programming languages, and $f_{out}$ can be computed for any language that supports controlled execution with comparable test cases. For cross-language clone detection, the same principle holds. Embeddings provide a shared representation space, and agreement at the output level serves as a language-agnostic semantic check.

\subsection{Limitations and Threats to Validity}

Output agreement depends on test suite adequacy. Finite input sets may overestimate equivalence between non-identical programs or miss rare divergences. This limitation applies to all execution-based similarity methods and does not arise from the proposed fusion architecture. Dropout applied to the fused representation and the use of a low-dimensional feature reduce over-reliance. Precision, recall, and f-measure agree with benchmark labels rather than formal semantic equivalence. In addition, $f{out}$ reflects behavior observed on a finite test set and does not constitute a proof of equivalence.

The evaluation relies on relatively short code fragments and controlled benchmarks, which may not reflect industrial settings that involve long functions, complex dependencies, and partial executability. GraphCodeBERT enforces a token length limit, and longer fragments may require chunking or hierarchical encoding. The feasibility and usefulness of $f_{out}$ also depend on the availability of safe execution. These threats clarify the conditions under which the method applies and the requirements for transfer to new corpora and deployment environments.

\section{Conclusion}

This study augments GraphCodeBERT with an execution-derived behavioral signature and a trainable fusion head. The resulting design provides a practical template for integrating external knowledge sources into pre-trained code models, with applications beyond clone detection in knowledge-driven software engineering tasks. The proposed model fuses the pooled transformer representation with a scalar feature computed from output agreement on a fixed test suite. The feature is projected into the hidden space and concatenated with the pooled representation, producing a richer input for the classifier. Experimental results show consistent gains over the baseline, supporting the claim that explicit behavioral evidence carries information that cannot be reliably inferred from tokens and structural signals alone.

Preliminary experiments on small subsets of BigCloneBench show behavior consistent with results observed on IR-Plag, indicating that the fusion mechanism extends beyond the primary benchmark. These additional evaluations are qualitative in nature, yet they align with the expected contribution of execution-derived evidence toward reducing remaining semantic errors.

Future work will assess the method on larger and more diverse code corpora, examine higher-dimensional behavioral signals, and investigate static alternatives to $f_{out}$ in settings where execution is unsafe. Additional directions include expanded test generation strategies aimed at increasing behavioral coverage.

\section*{Acknowledgments}
The research reported in this paper has been funded by the Federal Ministry for Climate Action, Environment, Energy, Mobility, Innovation, and Technology (BMK), the Federal Ministry for Digital and Economic Affairs (BMDW), and the State of Upper Austria in the frame of SCCH, a center in the COMET - Competence Centers for Excellent Technologies Programme.

\bibliography{mybib}

\begin{thebibliography}{23}
\expandafter\ifx\csname natexlab\endcsname\relax\def\natexlab#1{#1}\fi
\providecommand{\url}[1]{\texttt{#1}}
\providecommand{\href}[2]{#2}
\providecommand{\path}[1]{#1}
\providecommand{\DOIprefix}{doi:}
\providecommand{\ArXivprefix}{arXiv:}
\providecommand{\URLprefix}{URL: }
\providecommand{\Pubmedprefix}{pmid:}
\providecommand{\doi}[1]{\href{http://dx.doi.org/#1}{\path{#1}}}
\providecommand{\Pubmed}[1]{\href{pmid:#1}{\path{#1}}}
\providecommand{\bibinfo}[2]{#2}
\ifx\xfnm\relax \def\xfnm[#1]{\unskip,\space#1}\fi
\bibitem[{Ain et~al.(2019)Ain, Butt, Anwar, Azam \&
  Maqbool}]{ain2019systematic}
\bibinfo{author}{Ain, Q.~U.}, \bibinfo{author}{Butt, W.~H.},
  \bibinfo{author}{Anwar, M.~W.}, \bibinfo{author}{Azam, F.}, \&
  \bibinfo{author}{Maqbool, B.} (\bibinfo{year}{2019}).
\newblock \bibinfo{title}{A systematic review on code clone detection}.
\newblock {\it \bibinfo{journal}{IEEE access}\/},  {\it \bibinfo{volume}{7}\/},
  \bibinfo{pages}{86121--86144}.
\bibitem[{Alon et~al.(2019)Alon, Zilberstein, Levy \& Yahav}]{alon2019code2vec}
\bibinfo{author}{Alon, U.}, \bibinfo{author}{Zilberstein, M.},
  \bibinfo{author}{Levy, O.}, \& \bibinfo{author}{Yahav, E.}
  (\bibinfo{year}{2019}).
\newblock \bibinfo{title}{code2vec: Learning distributed representations of
  code}.
\newblock {\it \bibinfo{journal}{Proceedings of the ACM on Programming
  Languages}\/},  {\it \bibinfo{volume}{3}\/}, \bibinfo{pages}{1--29}.
\bibitem[{Feng et~al.(2020)Feng, Guo, Tang, Duan, Feng, Gong, Shou, Qin, Liu,
  Jiang \& Zhou}]{key-codebert}
\bibinfo{author}{Feng, Z.}, \bibinfo{author}{Guo, D.}, \bibinfo{author}{Tang,
  D.}, \bibinfo{author}{Duan, N.}, \bibinfo{author}{Feng, X.},
  \bibinfo{author}{Gong, M.}, \bibinfo{author}{Shou, L.}, \bibinfo{author}{Qin,
  B.}, \bibinfo{author}{Liu, T.}, \bibinfo{author}{Jiang, D.}, \&
  \bibinfo{author}{Zhou, M.} (\bibinfo{year}{2020}).
\newblock \bibinfo{title}{Codebert: {A} pre-trained model for programming and
  natural languages}.
\newblock In \bibinfo{editor}{T.~Cohn}, \bibinfo{editor}{Y.~He}, \&
  \bibinfo{editor}{Y.~Liu} (Eds.), {\it \bibinfo{booktitle}{Findings of the
  Association for Computational Linguistics: {EMNLP} 2020, Online Event, 16-20
  November 2020}\/} (pp. \bibinfo{pages}{1536--1547}).
\newblock \bibinfo{publisher}{Association for Computational Linguistics} volume
  \bibinfo{volume}{{EMNLP} 2020} of {\it \bibinfo{series}{Findings of
  {ACL}}\/}.
\bibitem[{Gabel et~al.(2008)Gabel, Jiang \& Su}]{gabel2008scalable}
\bibinfo{author}{Gabel, M.}, \bibinfo{author}{Jiang, L.}, \&
  \bibinfo{author}{Su, Z.} (\bibinfo{year}{2008}).
\newblock \bibinfo{title}{Scalable detection of semantic clones}.
\newblock In {\it \bibinfo{booktitle}{Proceedings of the 30th international
  conference on Software engineering}\/} (pp. \bibinfo{pages}{321--330}).
\bibitem[{Guo et~al.(2021)Guo, Ren, Lu, Feng, Tang, Liu, Zhou, Duan,
  Svyatkovskiy, Fu, Tufano, Deng, Clement, Drain, Sundaresan, Yin, Jiang \&
  Zhou}]{guo2020graphcodebert}
\bibinfo{author}{Guo, D.}, \bibinfo{author}{Ren, S.}, \bibinfo{author}{Lu, S.},
  \bibinfo{author}{Feng, Z.}, \bibinfo{author}{Tang, D.}, \bibinfo{author}{Liu,
  S.}, \bibinfo{author}{Zhou, L.}, \bibinfo{author}{Duan, N.},
  \bibinfo{author}{Svyatkovskiy, A.}, \bibinfo{author}{Fu, S.},
  \bibinfo{author}{Tufano, M.}, \bibinfo{author}{Deng, S.~K.},
  \bibinfo{author}{Clement, C.~B.}, \bibinfo{author}{Drain, D.},
  \bibinfo{author}{Sundaresan, N.}, \bibinfo{author}{Yin, J.},
  \bibinfo{author}{Jiang, D.}, \& \bibinfo{author}{Zhou, M.}
  (\bibinfo{year}{2021}).
\newblock \bibinfo{title}{Graphcodebert: Pre-training code representations with
  data flow}.
\newblock In {\it \bibinfo{booktitle}{9th International Conference on Learning
  Representations, {ICLR} 2021, Virtual Event, Austria, May 3-7, 2021}\/}.
\newblock \bibinfo{publisher}{OpenReview.net}.
\bibitem[{Hartanto et~al.(2019)Hartanto, Syaputra \&
  Pristyanto}]{hartanto2019best}
\bibinfo{author}{Hartanto, A.~D.}, \bibinfo{author}{Syaputra, A.}, \&
  \bibinfo{author}{Pristyanto, Y.} (\bibinfo{year}{2019}).
\newblock \bibinfo{title}{Best parameter selection of rabin-karp algorithm in
  detecting document similarity}.
\newblock In {\it \bibinfo{booktitle}{2019 International Conference on
  Information and Communications Technology (ICOIACT)}\/} (pp.
  \bibinfo{pages}{457--461}).
\newblock \bibinfo{organization}{IEEE}.
\bibitem[{Karmakar \& Robbes(2021)}]{karmakar2021pre}
\bibinfo{author}{Karmakar, A.}, \& \bibinfo{author}{Robbes, R.}
  (\bibinfo{year}{2021}).
\newblock \bibinfo{title}{What do pre-trained code models know about code?}
\newblock In {\it \bibinfo{booktitle}{2021 36th IEEE/ACM International
  Conference on Automated Software Engineering (ASE)}\/} (pp.
  \bibinfo{pages}{1332--1336}).
\newblock \bibinfo{organization}{IEEE}.
\bibitem[{Karnalim et~al.(2019)Karnalim, Budi, Toba \& Joy}]{key-karnalim}
\bibinfo{author}{Karnalim, O.}, \bibinfo{author}{Budi, S.},
  \bibinfo{author}{Toba, H.}, \& \bibinfo{author}{Joy, M.}
  (\bibinfo{year}{2019}).
\newblock \bibinfo{title}{Source code plagiarism detection in academia with
  information retrieval: Dataset and the observation.}
\newblock {\it \bibinfo{journal}{Informatics in Education}\/},  {\it
  \bibinfo{volume}{18}\/}, \bibinfo{pages}{321--344}.
\bibitem[{Karnalim \& Simon(2020)}]{karnalim2020syntax}
\bibinfo{author}{Karnalim, O.}, \& \bibinfo{author}{Simon}
  (\bibinfo{year}{2020}).
\newblock \bibinfo{title}{Syntax trees and information retrieval to improve
  code similarity detection}.
\newblock In {\it \bibinfo{booktitle}{Proceedings of the Twenty-Second
  Australasian Computing Education Conference}\/} (pp.
  \bibinfo{pages}{48--55}).
\bibitem[{Karnalim et~al.(2021)}]{karnalim2021explanation}
\bibinfo{author}{Karnalim, O.} et~al. (\bibinfo{year}{2021}).
\newblock \bibinfo{title}{Explanation in code similarity investigation}.
\newblock {\it \bibinfo{journal}{IEEE Access}\/},  {\it \bibinfo{volume}{9}\/},
  \bibinfo{pages}{59935--59948}.
\bibitem[{Martinez-Gil(2022)}]{key-martinez-mlwa}
\bibinfo{author}{Martinez-Gil, J.} (\bibinfo{year}{2022}).
\newblock \bibinfo{title}{A comprehensive review of stacking methods for
  semantic similarity measurement}.
\newblock {\it \bibinfo{journal}{Machine Learning with Applications}\/},  {\it
  \bibinfo{volume}{10}\/}, \bibinfo{pages}{100423}.
\bibitem[{Martinez-Gil(2023)}]{key-martinez-ijseke}
\bibinfo{author}{Martinez-Gil, J.} (\bibinfo{year}{2023}).
\newblock \bibinfo{title}{A comparative study of ensemble techniques based on
  genetic programming: {A} case study in semantic similarity assessment}.
\newblock {\it \bibinfo{journal}{Int. J. Softw. Eng. Knowl. Eng.}\/},  {\it
  \bibinfo{volume}{33}\/}, \bibinfo{pages}{289--312}.
\bibitem[{Martinez-Gil(2024)}]{key-martinez-swqd}
\bibinfo{author}{Martinez-Gil, J.} (\bibinfo{year}{2024}).
\newblock \bibinfo{title}{Source code clone detection using unsupervised
  similarity measures}.
\newblock In \bibinfo{editor}{P.~Bludau}, \bibinfo{editor}{R.~Ramler},
  \bibinfo{editor}{D.~Winkler}, \& \bibinfo{editor}{J.~Bergsmann} (Eds.), {\it
  \bibinfo{booktitle}{Software Quality as a Foundation for Security - 16th
  International Conference on Software Quality, {SWQD} 2024, Vienna, Austria,
  April 23-25, 2024, Proceedings}\/} (pp. \bibinfo{pages}{21--37}).
\newblock \bibinfo{publisher}{Springer} volume \bibinfo{volume}{505} of {\it
  \bibinfo{series}{LNBIP}\/}.
\bibitem[{Martinez-Gil(2025{\natexlab{a}})}]{key-martinez-codesim}
\bibinfo{author}{Martinez-Gil, J.} (\bibinfo{year}{2025}{\natexlab{a}}).
\newblock \bibinfo{title}{Advanced detection of source code clones via an
  ensemble of unsupervised similarity measures}.
\newblock In \bibinfo{editor}{J.~Fischbach}, \bibinfo{editor}{R.~Ramler},
  \bibinfo{editor}{D.~Winkler}, \& \bibinfo{editor}{J.~Bergsmann} (Eds.), {\it
  \bibinfo{booktitle}{Balancing Software Innovation and Regulatory Compliance -
  17th International Conference on Software Quality, {SWQD} 2025, Munich,
  Germany, May 20-22, 2025, Proceedings}\/} (pp. \bibinfo{pages}{72--90}).
\newblock \bibinfo{publisher}{Springer} volume \bibinfo{volume}{544} of {\it
  \bibinfo{series}{Lecture Notes in Business Information Processing}\/}.
\bibitem[{Martinez-Gil(2025{\natexlab{b}})}]{key-martinez-ijseke2}
\bibinfo{author}{Martinez-Gil, J.} (\bibinfo{year}{2025}{\natexlab{b}}).
\newblock \bibinfo{title}{Augmenting the interpretability of graphcodebert for
  code similarity tasks}.
\newblock {\it \bibinfo{journal}{Int. J. Softw. Eng. Knowl. Eng.}\/},  {\it
  \bibinfo{volume}{35}\/}, \bibinfo{pages}{657--678}. \URLprefix
  \url{https://doi.org/10.1142/S0218194025500160}.
  \DOIprefix\doi{10.1142/S0218194025500160}.
\bibitem[{Martinez{-}Gil(2025)}]{key-martinez-eswa}
\bibinfo{author}{Martinez{-}Gil, J.} (\bibinfo{year}{2025}).
\newblock \bibinfo{title}{Framework to automatically determine the quality of
  open data catalogs}.
\newblock {\it \bibinfo{journal}{Expert Syst. Appl.}\/},  {\it
  \bibinfo{volume}{289}\/}, \bibinfo{pages}{128379}. \URLprefix
  \url{https://doi.org/10.1016/j.eswa.2025.128379}.
  \DOIprefix\doi{10.1016/J.ESWA.2025.128379}.
\bibitem[{Mou et~al.(2016)Mou, Li, Zhang, Wang \& Jin}]{mou2016convolutional}
\bibinfo{author}{Mou, L.}, \bibinfo{author}{Li, G.}, \bibinfo{author}{Zhang,
  L.}, \bibinfo{author}{Wang, T.}, \& \bibinfo{author}{Jin, Z.}
  (\bibinfo{year}{2016}).
\newblock \bibinfo{title}{Convolutional neural networks over tree structures
  for programming language processing}.
\newblock In {\it \bibinfo{booktitle}{Proceedings of the AAAI conference on
  artificial intelligence}\/}.
\newblock volume~\bibinfo{volume}{30}.
\bibitem[{Novak et~al.(2019)Novak, Joy \& Kermek}]{novak2019source}
\bibinfo{author}{Novak, M.}, \bibinfo{author}{Joy, M.}, \&
  \bibinfo{author}{Kermek, D.} (\bibinfo{year}{2019}).
\newblock \bibinfo{title}{Source-code similarity detection and detection tools
  used in academia: a systematic review}.
\newblock {\it \bibinfo{journal}{ACM Transactions on Computing Education
  (TOCE)}\/},  {\it \bibinfo{volume}{19}\/}, \bibinfo{pages}{1--37}.
\bibitem[{Saini et~al.(2018)Saini, Singh et~al.}]{saini2018code}
\bibinfo{author}{Saini, N.}, \bibinfo{author}{Singh, S.} et~al.
  (\bibinfo{year}{2018}).
\newblock \bibinfo{title}{Code clones: Detection and management}.
\newblock {\it \bibinfo{journal}{Procedia computer science}\/},  {\it
  \bibinfo{volume}{132}\/}, \bibinfo{pages}{718--727}.
\bibitem[{Svajlenko \& Roy(2015)}]{svajlenko2015bigclonebench}
\bibinfo{author}{Svajlenko, J.}, \& \bibinfo{author}{Roy, C.~K.}
  (\bibinfo{year}{2015}).
\newblock \bibinfo{title}{Evaluating clone detection tools with bigclonebench}.
\newblock In {\it \bibinfo{booktitle}{2015 IEEE international conference on
  software maintenance and evolution (ICSME)}\/} (pp.
  \bibinfo{pages}{131--140}).
\bibitem[{Wang et~al.(2020)Wang, Li, Ma, Xia \& Jin}]{wang2020detecting}
\bibinfo{author}{Wang, W.}, \bibinfo{author}{Li, G.}, \bibinfo{author}{Ma, B.},
  \bibinfo{author}{Xia, X.}, \& \bibinfo{author}{Jin, Z.}
  (\bibinfo{year}{2020}).
\newblock \bibinfo{title}{Detecting code clones with graph neural network and
  flow-augmented abstract syntax tree}.
\newblock In {\it \bibinfo{booktitle}{2020 IEEE 27th International Conference
  on Software Analysis, Evolution and Reengineering (SANER)}\/} (pp.
  \bibinfo{pages}{261--271}).
\newblock \bibinfo{organization}{IEEE}.
\bibitem[{Wei \& Li(2017)}]{wei2017supervised}
\bibinfo{author}{Wei, H.}, \& \bibinfo{author}{Li, M.} (\bibinfo{year}{2017}).
\newblock \bibinfo{title}{Supervised deep features for software functional
  clone detection by exploiting lexical and syntactical information in source
  code.}
\newblock In {\it \bibinfo{booktitle}{IJCAI}\/} (pp.
  \bibinfo{pages}{3034--3040}).
\bibitem[{White et~al.(2016)White, Tufano, Vendome \&
  Poshyvanyk}]{white2016deep}
\bibinfo{author}{White, M.}, \bibinfo{author}{Tufano, M.},
  \bibinfo{author}{Vendome, C.}, \& \bibinfo{author}{Poshyvanyk, D.}
  (\bibinfo{year}{2016}).
\newblock \bibinfo{title}{Deep learning code fragments for code clone
  detection}.
\newblock In {\it \bibinfo{booktitle}{Proceedings of the 31st IEEE/ACM
  international conference on automated software engineering}\/} (pp.
  \bibinfo{pages}{87--98}).

\end{thebibliography}

\end{document}